\documentstyle{mn}
\input psfig
\newif\ifAMStwofonts

\ifoldfss
  \ifCUPmtlplainloaded \else
    \NewTextAlphabet{textbfit} {cmbxti10} {}
    \NewTextAlphabet{textbfss} {cmssbx10} {}
    \NewMathAlphabet{mathbfit} {cmbxti10} {} 
    \NewMathAlphabet{mathbfss} {cmssbx10} {} 
  \fi
  \ifAMStwofonts
    \ifCUPmtlplainloaded \else
      \NewSymbolFont{upmath} {eurm10}
      \NewSymbolFont{AMSa} {msam10}
      \NewMathSymbol{\upi}     {0}{upmath}{19}
      \NewMathSymbol{\umu}     {0}{upmath}{16}
      \NewMathSymbol{\upartial}{0}{upmath}{40}
      \NewMathSymbol{\leqslant}{3}{AMSa}{36}
      \NewMathSymbol{\geqslant}{3}{AMSa}{3E}

    \fi
  \fi
\fi 

\ifnfssone
  \newmathalphabet{\mathit}
  \addtoversion{normal}{\mathit}{cmr}{m}{it}
  \addtoversion{bold}{\mathit}{cmr}{bx}{it}
  \newmathalphabet{\mathbfit} 
  \addtoversion{normal}{\mathbfit}{cmr}{bx}{it}
  \addtoversion{bold}{\mathbfit}{cmr}{bx}{it}
  \newmathalphabet{\mathbfss} 
  \addtoversion{normal}{\mathbfss}{cmss}{bx}{n}
  \addtoversion{bold}{\mathbfss}{cmss}{bx}{n}
  \ifAMStwofonts
    \ifCUPmtlplainloaded \else
      \UseAMStwoboldmath
      \new@mathgroup\upmath@group
      \define@mathgroup\mv@normal\upmath@group{eur}{m}{n}
      \define@mathgroup\mv@bold\upmath@group{eur}{b}{n}
      \edef\UPM{\hexnumber\upmath@group}
      \new@mathgroup\amsa@group
      \define@mathgroup\mv@normal\amsa@group{msa}{m}{n}
      \define@mathgroup\mv@bold\amsa@group{msa}{m}{n}
      \edef\AMSa{\hexnumber\amsa@group}
      \makeatother
      \mathchardef\upi="0\UPM19
      \mathchardef\umu="0\UPM16
      \mathchardef\upartial="0\UPM40
      \mathchardef\leqslant="3\AMSa36
      \mathchardef\geqslant="3\AMSa3E
    \fi
  \fi
\fi 

\ifnfsstwo
  \DeclareMathAlphabet{\mathbfit}{OT1}{cmr}{bx}{it}
  \SetMathAlphabet\mathbfit{bold}{OT1}{cmr}{bx}{it}
  \DeclareMathAlphabet{\mathbfss}{OT1}{cmss}{bx}{n}
  \SetMathAlphabet\mathbfss{bold}{OT1}{cmss}{bx}{n}
  \ifAMStwofonts
    \ifCUPmtlplainloaded \else
      \DeclareSymbolFont{UPM}{U}{eur}{m}{n}
      \SetSymbolFont{UPM}{bold}{U}{eur}{b}{n}
      \DeclareSymbolFont{AMSa}{U}{msa}{m}{n}
      \DeclareMathSymbol{\upi}{0}{UPM}{"19}
      \DeclareMathSymbol{\umu}{0}{UPM}{"16}
      \DeclareMathSymbol{\upartial}{0}{UPM}{"40}
      \DeclareMathSymbol{\leqslant}{3}{AMSa}{"36}
      \DeclareMathSymbol{\geqslant}{3}{AMSa}{"3E}
    \fi
  \fi
\fi 

\ifCUPmtlplainloaded \else
  \ifAMStwofonts \else 
    \def\upi{\pi}
    \def\umu{\mu}
    \def\upartial{\partial}
  \fi
\fi

\title[{\it ROSAT} PSPC observation towards PC1643+4631: no cluster]
{A deep {\it ROSAT} PSPC observation towards the CMB decrement close to 
PC1643+4631 A \& B: no cluster X-ray emission}

\author[Kneissl, Sunyaev and White]
  {R\"udiger~Kneissl, Rashid~A.~Sunyaev and Simon~D.M.~White\\
Max-Planck-Institut f\"ur Astrophysik, Karl-Schwarzschild-Str.1,
85740 Garching, Germany}
\date{Accepted Received 17 December 1997}
\pagerange{\pageref{firstpage}--\pageref{lastpage}}
\pubyear{}

\def\LaTeX{L\kern-.36em\raise.3ex\hbox{a}\kern-.15em
    T\kern-.1667em\lower.7ex\hbox{E}\kern-.125emX}

\begin{document}

\label{firstpage}

\maketitle

\begin{abstract}
We report on a 16 ks observation with the {\it ROSAT} PSPC
centred on the CMB decrement which Jones et al. (1997) report
in the direction of the quasar pair PC1643+4631. We do not detect 
the X-ray emission which would be expected if the decrement were
caused by Compton up-scattering of CMB photons by hot gas.  Our upper 
limit on the bolometric X-ray flux within a circle of radius 1\farcm 5 is
$1.9 \times 10^{-14}$ ergs cm$^{-2}$ s$^{-1}$ at 99.7\% significance.
This is based on data in the energy  range 0.5-2.0 keV and assumes
bremsstrahlung with an observed temperature of 2.5 keV.
We investigate the requirements for a spherical body of hot gas
to produce the decrement while remaining consistent with our
data. With conservative assumptions we establish a lower redshift limit
of 2.8 at 95 \% confidence for a 10 keV isothermal object of any
size. Even higher limits obtain for objects with temperatures 
between 0.2 and 10 keV. We conclude that the CMB decrement is 
unlikely to be caused by a galaxy cluster. 
We discuss alternatives to a single, spherical cluster, 
such as a double or elongated cluster, a supercluster/filament, 
or scattering by a moving cluster. None of these appears 
attractive, especially since another candidate for a decrement near a 
QSO pair and without an obvious associated cluster has been found. 
An explanation other than Comptonization by hot gas within 
a virialised object appears needed.
\end{abstract}


\section{Introduction}

In an intriguing paper Jones et al. \shortcite{jon97} report the discovery 
of a decrement in the cosmic microwave background (CMB) in the direction of
the quasar pair PC1643+4631 A \& B. They suggest this may be due 
to a distant galaxy cluster which might also produce the
quasar pair  through gravitational lensing of a single source.  The
lack of detected X-ray emission in this field enabled Jones et al 
to establish that any such cluster must be at a redshift of 1 or
greater. In this paper we report results from a pointed {\it ROSAT} 
observation which improves this X-ray constraint by a factor of about 20.

The interferometric data of Jones et al were taken with the Ryle
Telescope (RT) at 15 GHz and give a flux decrement of $-380 \pm 64 \mu$Jy 
averaged over the $110'' \times 175''$ beam. For a circularly symmetric 
signal, this corresponds to a true decrement of at least 560 $\mu$K,
this minimum value implying a core radius of about
$60''$, or roughly $250 h^{-1}$ kpc at high 
redshift. (Here $h= H_0/(100{\rm km/s/Mpc})$.) Since intrinsic
fluctuations in the CMB are not expected on arc-minute scales, the
most plausible origin for this decrement
is inverse Compton scattering of CMB photons by hot
intracluster gas (Sunyaev \& Zel'dovich 1972). Decrements of this size
have been detected in the direction of a number of massive galaxy
clusters (Birkinshaw, Gull, Hardebeck 1984; Birkinshaw, Hughes, 
Arnaud 1991; Jones et al. 1993; Grainge et al. 1993; Birkinshaw, 
Hughes 1994; Wilbanks et al. 1994; Herbig et al. 1995; Grainge et
al. 1996; Carlstrom, Joy, Grego 1997; Holzapfel et al. 1997; Myers
et al. 1997). As Jones et al. discuss, such intracluster gas
should be detectable in X-rays. They were able to put an upper limit
on the X-ray flux using a 11.4 ks observation with the {\it ROSAT}
PSPC in which the quasar pair serendipitously lies just within the 
field of view. Unfortunately vignetting and smearing by the point
spread function greatly reduce the sensitivity of this observation,
and we were able to establish a comparable (but independent) upper
limit using data from {\it ROSAT} All-Sky Survey, which has a local 
exposure time of only 801 sec. Both observations limit the flux to be
below $2 \times 10^{-13}$ erg cm$^{-2}$ s$^{-1}$ (0.1-2.4 keV) at 
99.7\% confidence. Jones et al show that this constrains 
any cluster to be at $z>1$. An extensive search has so far identified 
no optical or infrared counterpart \cite{sau97a}. An HST observation 
has been scheduled (7342 by Saunders et al.) which should probe 
to even deeper limits.

If the observed decrement is indeed due to Compton scattering from 
hot gas, then X-ray observations offer the only way to detect this gas 
directly. The {\it ROSAT} detectors, with an energy range of 
0.1--2.4 keV, are ideally suited to observe high redshift 
clusters which are expected to have temperatures in the range 1--10 keV. 
We therefore proposed and obtained a long pointed PSPC observation 
of this field. 

\section{The {\it ROSAT} PSPC observation}

The observation, carried out between the 25th of February and the 3rd
of March, was centred on RA = 16$^{\rm h}$45$^{\rm m}$12$^{\rm s}$, 
DEC = +46\degr24$'$35$''$ (J 2000), and had a total accepted exposure time
of 15889 s. This was among the very last observations with the 
PSPC before complete exhaustion of its gas supply. A small ($\sim$ 10 
\%) drop in the gain occurred towards the end of our observation period 
and has been taken into account in the following analysis. 
Only few of the 61 X-ray sources detected in the observation have so 
far been identified, and all lie more than 35$'$ off-axis. We 
were, nevertheless, able to check the absolute positioning of the 
{\it ROSAT} XRT by comparing several sources within the support
structure 
with objects on the Palomar plates. This confirmed our estimated 
coordinates to be accurate to a few arcseconds. Our astrometry is thus
much better than the width of the psf. 
\begin{figure}
\psfig{file=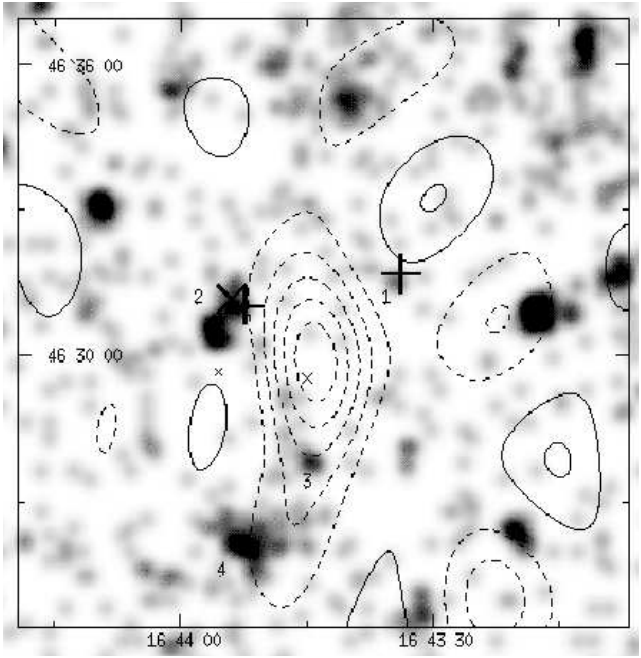,width=\hsize,%
bbllx=82pt,bblly=165pt,bburx=528pt,bbury=628pt,clip=}
\caption{Overlay of the CMB measurement \protect\cite{jon97} 
on our X-ray observation; coordinates in this plot are B1950.
The separation of the quasar pair (+) PC1643+4631 A (right) \& B
(left) is 198$''$. Positions of removed 
radio sources ($\times$) 
are also marked. X-ray sources (see Tab.~\ref{tab:sources}) are 
labelled 1--4.} 
\label{fig:overlay}
\end{figure}
\begin{table*}
\begin{minipage}{120mm}
\caption{Source detections with likelihood $>$ 5 near 
the centre of the decrement.}
\begin{tabular}{ccccccccc}
Object & RA (251\degr+) & DEC (46\degr+) & ML & Dist & Cts & Ext & Ext ML & Rate \\
 & & & & & & & & \\
1 & .25658 & .43541 & 6 & 132 & 4.9 $\pm$ 2.5 & & & 3.3 $\pm$ 1.7 \\
2 & .34073 & .42180 & 73 & 122 & 43.8 $\pm$ 7.0 & 35 & 8 & 29.3 $\pm$
4.7 \\
3 & .29930 & .37334 & 18 & 152 & 10.6 $\pm$ 3.6 & & & 7.1 $\pm$ 2.4 \\
4 & .32700 & .34584 & 78 & 263 & 64.1 $\pm$ 8.6 & 57 & 30 & 43.1 $\pm$
5.8 \\
 & & & & & & & & \\
\multicolumn{9}{l}{{\footnotesize ML: Likelihood ratio, Dist:
distance to the centre of the CMB decrement in arcsec}} \\
\multicolumn{9}{l}{{\footnotesize Cts: counts, Ext: extent (FWHM in
arcsec), Rate: count rate in $10^{-4}$ ct s$^{-1}$}} \\
\end{tabular}
\label{tab:sources}
\end{minipage}
\end{table*}

Fig.~\ref{fig:overlay} shows an overlay of the CMB measurement on our
X-ray map (in the 0.5 -- 2.0 keV band). 
A pixelization 20 times finer than the local psf 
(which has FWHM of 37\farcs6) has been chosen, followed by 
Gaussian smoothing with a third of the width of the psf. The faintest
grey patches are individual photons. We have labelled four sources
near the decrement with a detection likelihood ratio exceeding 5, and
we give their properties in Table 1. This very low detection threshold
was chosen to demonstrate that not even a marginally detected source lies 
close to the centre of the microwave decrement. Secure detections are 
commonly quoted for a likelihood ratio of at least 10 ($\sim$ 4-$\sigma$). 
Objects 2 and 4 are both significantly more extended than the psf. 
Visual inspection suggests that 2 could consist of two close point
sources, while 4 might really be an extended object. There are too few
counts to use spectral analysis to check the possibility of
a galaxy cluster. Other data \cite{bru96} suggest that X-ray
emission from the quasars might be detectable in our data. We find no
secure detection at the quasar positions, but there are 
marginal detections close to both of them (source 
1 and possibly one component (with roughly 
half the total count rate) of the extended source 
2). We use these to set 2$\sigma$ upper limits of 7 and 
$20 \times 10^{-4}$ ct s$^{-1}$ on sources associated with A and B 
respectively. For typical X-ray energy indices of $\alpha_x$ = 0.5--1.0 the 
monochromatic 
luminosities are 7.5 and $21 \times 10^{44}$ erg s$^{-1}$ keV$^{-1}$, and 
with the optical magnitudes of $M_B = -26.1$ and -25.8 
(Schneider, Schmidt, Gunn 1991) we find limits on the optical to X-ray
spectral indices of $\alpha_{ox} > 1.4$ and 1.2. 

The most stringent and robust upper limits on cluster X-ray emission
associated with the decrement are obtained using the observed energy 
interval 0.5-2.0 keV, since the Galactic background is high at lower 
energies and the instrument has little sensitivity at higher 
energies. Results from including the 0.1-0.5 keV band, more stringent 
for intrinsic gas temperatures below 1 keV, 
are discussed later in the paper. 
We quote limits within a circular aperture of 
radius 1\farcm5  centred on RA = 251\fdg29583, DEC = +46\fdg40687, 
the centre of the decrement. This extends over the 
3 $\sigma$ contours of the microwave observation. 
Within this area we have 13 counts compared to the expected background
of 15 counts. (Consistent background estimates were obtained by a
spline fit to the image after removal of detected sources, and by averaging
random regions without obvious sources and more than 10$'$ from the
decrement.) With a Gaussian approximation to the Poisson photon arrival
distribution we find that a source with an expected count of 12.5
can be excluded with 99.7 \% confidence. This 
corresponds to a limiting count rate of $8.9 \times 10^{-4}$ cts s$^{-1}$. 
Deconvolving the effective area and response of the instrument, 
and correcting for Galactic absorption due to an HI column density of 
$1.8 \times 10^{20}$ cm$^{-2}$, this corresponds to 
a bolometric flux limit of $1.9 \times 10^{-14}$ erg cm$^{-2}$ 
s$^{-1}$ for an assumed bremsstrahlung spectrum with 
temperature 2.5 keV in the observer frame. This upper limit is almost 
twenty times lower than that found using the {\it ROSAT} Sky Survey
data or the serendipitous observation analysed by Jones et al. 

\section{Cluster Models}

\begin{figure}
\psfig{file=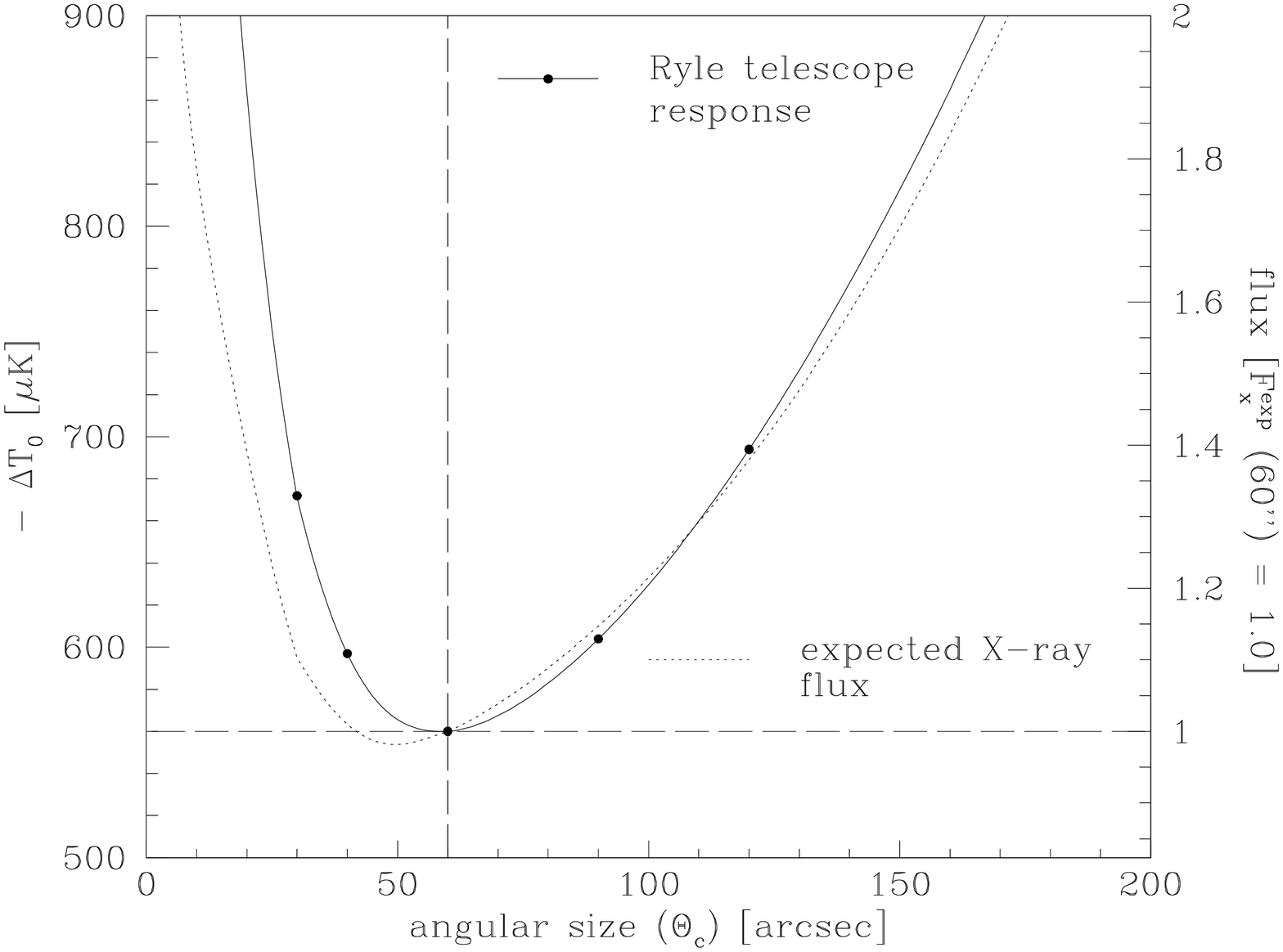,width=\hsize,clip=}
\caption{The amplitude of the true central temperature decrement
(solid line) and the expected X-ray flux (in arbitrary units, dotted
line) is plotted against the core radius for an
isothermal model with density profile given by equ.~\ref{eq:beta}.
These data (points) were kindly provided by the Cambridge MRAO group.}
\label{fig:rtresp}
\end{figure}

We now study the constraints imposed on a possible high redshift
cluster by the combined X-ray and radio data. We follow
Jones et al. in modelling the intracluster medium as an isothermal
gas, temperature $T$, distributed spherically with density profile
\begin{equation}
\label{eq:beta}
n_e (r) \, =  \, n_0 \, (1+r^2/r_c^2)^{-1}.
\end{equation}
To determine the central value of the temperature decrement 
a detailed knowledge of the synthesised beam, including the scanning 
path, is required. In 
practise the response of the telescope to different profiles must be 
modelled and fit directly to the visibility data. 
Jones et al. fit various core radii for 
the model of equ.~\ref{eq:beta}, which implies a 
temperature decrement with profile 
\begin{equation}
\label{eq:betaproj}
\Delta T = \Delta T_0 (1 + \Theta^2 / 
\Theta_c^2)^{-\frac{1}{2}}.
\end{equation}
The results were kindly
provided to us by Richard Saunders and are shown in Fig.~\ref{fig:rtresp}.
For this model an X-ray flux limit $F_{x}^{lim}$ within an aperture
of radius $\Theta_a$ implies 
\begin{eqnarray}
\lefteqn{(1+z)^{4} \, r_A (H_0,\Omega,z) \,\,
\frac{(kT)^2}{f(T)} \,\,\,\, > } \nonumber\\
 & & \frac{1}{4 \pi} \, 
\left(\frac{m_e c^2}{\sigma_T} \right)^2 \, \frac{\Theta_c
y_0^2}{F_{x}^{lim}} \, \left( 1 - \left[1 + \Theta_a^2 / \Theta_c^2 \right]^{-\frac{1}{2}} \right), 
\end{eqnarray}
where 
$f(T) = {\cal L}_x / n_e^2$ is the 
X-ray cooling function, $y_0$ the central decrement, $\Theta_c$ the
angular core radius of the cluster model, and $r_A (H_0,\Omega,z)$ 
is the angular-size distance.
All quantities on the right
of this inequality are directly observable, although in practice the
radio data only constrain a combination of $y_0$ and $\Theta_c$
(Fig.~\ref{fig:rtresp}). It turns out that a core radius near
$1^\prime$ not only yields the minimum true central temperature decrement, 
but also the lowest value of the right-hand side of 
equation 2 (for given $F_{x}^{lim}$). It thus gives
the most conservative lower limit on the redshift of the cluster. 
Given the parameter dependence of the left-hand side, we will clearly
get the lowest possible limit on $(1+z)$ if we take the largest
plausible values for $T$ and $r_A$. We will quote limits mostly
for $T=10$~keV, $h=0.5$ and $\Omega=1$. This choice for $\Omega$ 
is not conservative, but we will show below that our limits are only
slightly reduced even for relatively extreme choices of the cosmological
model.

The RT map is made for a very limited set of interferometer spacings, 
and hence
only constrains the structure of the decrement on a small range of
angular scales. As a result, interpretation of the radio data is quite
model-dependent. 
The telescope response to model profiles other that of
equ.~\ref{eq:betaproj} has not been 
studied, so we present our limits for this case only. Since $\Theta_c$
is comparable to our X-ray resolution and also to the aperture we use
to put limits on the X-ray flux, we do not expect a significant
weakening of our limits for other reasonable
profiles. (Equ.~\ref{eq:beta} 
is, in fact, a relatively good fit to most clusters studied so far.)
We stress here that most plausible deviations from this model, for 
example a cooling flow in the inner regions, a temperature drop 
at large radii, 
or substructure in the gas, would actually enhance the X-ray 
emission relative to the microwave decrement, and thus would 
strengthen our limits. 

Table 2 gives characteristic  parameters for the gas halos at 
redshifts 1.25 and 3
which would produce the minimal X-ray emission consistent with 
the RT data; in both cases we assume $T=10$ keV, $h=0.5$ and
$\Omega=1$. For both redshifts the derived parameters
are within the range observed for nearby clusters, although the 
luminosities are somewhat low and the core radii somewhat large
as a result of our conservative assumptions.
Table 2 also lists the upper limit on
the X-ray luminosity implied by our PSPC observation. Clearly a cluster at
$z=1.25$ is strongly excluded, while a cluster at $z=3$ is marginally
consistent with our data.
\begin{table}
\caption{Gas halo parameters at $z \sim 1$ and 3 for T = 10 keV and
the model of equ.~\ref{eq:beta}}
\begin{tabular}{llll}
{\bf z} & {\bf 1.25} & {\bf 3} & \\
Core radius & 517 & 436 & kpc\\
Central electron density & 1.6 & 1.9 & $\times 10^{-3}$ cm$^{-3}$ \\
Emission integral & 3.3 & 4.7 & $\times 10^{-6}$ cm$^{-6}$ Mpc$^{3}$ \\
X-ray luminosity (bol.) & 2.0 & 2.8 & $\times 10^{45}$ erg s$^{-1}$ \\
X-ray limit (2-$\sigma$, bol.) & 0.3 & 3.4 & $\times 10^{45}$ erg s$^{-1}$ \\
\end{tabular}
\end{table}
\begin{figure}
\psfig{file=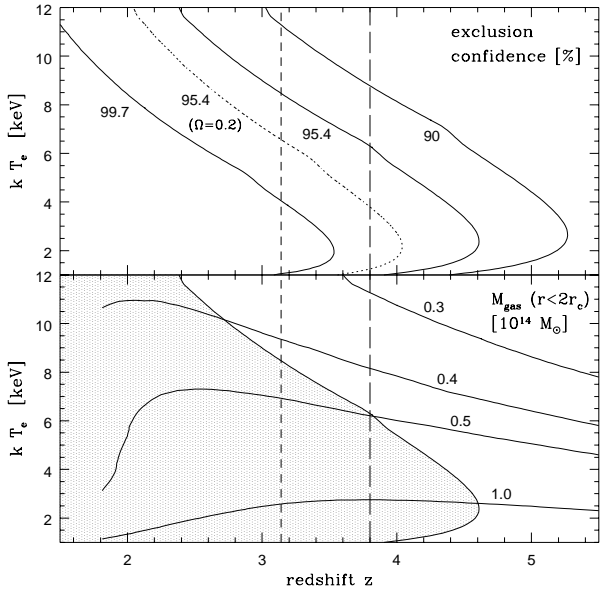,width=\hsize,clip=}
\caption{The upper panel shows confidence contours for excluding 
the hot gas model of (equ.~\ref{eq:beta}) based on the combined
radio and X-ray data. The gas masses which yield the most consistent
model at each redshift and temperature are shown in the lower panel 
(these are quoted in units of $10^{14} M_\odot$ and refer to the mass
within a sphere of radius corresponding to 2$'$). The heavy lines in 
the upper plot give confidence
levels for an $\Omega = 1$ universe, while the dotted line gives the 
95.4 \% level for an open universe with $\Omega = 0.2$. The long and
short dashed lines indicate the redshift of the quasar pair and of a damped
Ly$\alpha$ absorption feature in the spectrum of quasar
A. These plots are based on scalings derived from a
thermal Bremsstrahlung model which assumes the standard helium
abundance 
and no metals. We have checked the results against full spectral
modelling including heavy elements, Galactic foreground absorption,
and the response of the PSPC. Our results are little affected by these
factors for emitted temperatures above 2 keV, but at lower
temperatures they become sensitive to uncertainties in metallicity and
absorption. Results from detailed spectral modelling for the allowed range 
of low gas temperatures are discussed in the text.}
\label{fig:limits}
\end{figure}
To explore the limits imposed on redshift and gas temperature in a
more systematic way, we combine the radio and
the X-ray data
and maximise the joint likelihood of the two observations by varying 
the gas mass. Other model parameters are kept fixed, $h=0.5, r_c =
50''$ and $ \Omega = 1$ or 0.2. The results are presented 
in Fig.~\ref{fig:limits}. 
Hot halos consistent with both datasets are required
to lie at high redshift for any gas mass. 
For an assumed temperature of 5 keV,
a redshift of 4.2 is required to get consistency at the 95\%
confidence level. This is beyond the redshift of the quasar pair ($z$=3.8). 
For an even hotter halo with $T= 10$ keV, the redshift is still required to
exceed 2.8 at 95\% confidence. As can be seen by comparing the limits 
for an Einstein--de~Sitter universe and for an open universe with
$\Omega = 0.2$, these limits are quite insensitive to the cosmological
parameters. It is interesting that the hot cluster models which 
allow the lowest redshifts actually have rather small gas masses
compared to nearby clusters of comparable temperature.

\section{Discussion}

The X-ray flux predicted by our models could be reduced by a factor 
of 2 by assuming the cluster to be stretched along the
line-of-sight to twice its apparent diameter or to be broken into two
equal clusters which are superposed in projection. In fact, however, a
factor of two is not a great help, since it only results in a  
$\sim$ 25 \% reduction of our redshift limits. If the decrement is
due to a virialised cluster, it seems that we need a cluster at least 
as massive as Coma at a redshift beyond 3. The existence of even one 
such cluster in the observable universe is quite unexpected in
current structure formation models. When normalised to give the
presently observed abundance of clusters, only open models with 
$\Omega_0 < 0.25$ predict more than one cluster on the whole sky with 
$3<z<4$ and with $M>2\times 10^{15}M_\odot$. (This calculation is based
on a Press-Schechter model similar to those in White et al (1993).)
Such models overpredict the amplitude of fluctuations in the CMB. 

Gravitationally confined gas which is considerably hotter than 10 keV 
has not been found observationally on the spatial scales of interest. At low 
temperature on the other 
hand large-scale gas properties are somewhat uncertain. 
The lower temperature window indicated by the drop in redshift below 
2 keV in Fig.~\ref{fig:limits} might 
therefore be more promising for finding a plausible solution to the given 
data. Indeed the formal solution for a spherically symmetric 
object at redshift 3 and with temperature 0.2 keV 
could explain all observational data including the possible lensing for a 
gas mass fraction of 0.1. To derive this, the limit of $1.3 \times 10^{-3}$ 
ct s$^{-1}$ in the 0.1-2 keV energy interval was used, and very conservative 
assumptions about the plasma element abundances (zero metallicity, strongly 
suppressing line emission) and 
Galactic absorption ($N_H = 3.6 \times 10^{20}$ cm$^{-2}$, to account for 
local enhancements), which the 
limits in this temperature range depend sensitively upon, were made. 
The formation of such an object, however, would be hard to understand 
in the context of gravitational collapse. 
The implied gas mass of a few $\times 10^{15} M_\odot$ 
within a volume of $0.52 h^{-3}$ Mpc$^{3}$ 
has not been seen in any object and the assumed temperature is 
by two orders of magnitude lower then what would be expected for 
the virial temperature. 

Substantial reductions in X-ray flux are possible if we give
up the assumption that we are seeing a collapsed and virialised
system. The interferometer data require the extent of the object 
to be roughly $250 h^{-1}$ kpc in the directions transverse to the
line-of-sight, but its depth could be much greater; it would then
correspond to a supercluster ``filament'' seen end-on. Since such
filaments are not fully virialised they can have temperatures which
are substantially smaller than those of rich clusters, even though
their masses are comparable or even larger. The gas in such filaments
is heated by an accretion shock as it flows in from surrounding
regions and by local virialisation shocks within subclumps along the
filament. Typical velocities at these shocks are a few hundred km/s,
so that post-shock temperatures of a few million degrees are expected.
Thus at redshifts greater than 2 a prolate filament obeying equation
(1) with $r_c=50''$ but elongated by a factor of ten in depth 
(corresponding to a filament ``length'' of $\sim 5h^{-1}$Mpc) can be 
consistent with both X-ray and radio data for gas temperatures below
0.5 keV. The gas mass required is around $1 \times 10^{15}M_\odot$ (for 
$h=0.5$). This high gas mass and the {\it a priori} implausible aspect
of a close alignment of the filament with the line-of-sight 
disfavours this model, but careful quantitative analysis of 
numerical simulations is needed to see 
whether sufficiently massive filaments are formed at high
redshift in currently popular theories for structure formation. 

Gas in an aligned filament could also produce the observed microwave
decrement  without significant X-ray emission if it happened to be
flowing along the filament at high enough speed. The amplitude of
the observed decrement would be reproduced by Compton scattering from
$3\times 10^{14}M_\odot$ of gas moving away from us at 850~km/s,
regardless of the gas temperature. The speed required scales
inversely with the mass of gas, and, as before, the gas needs to
be projected onto a region with transverse diameter about
$0.5h^{-1}$Mpc.  Again numerical simulations are
required to estimate whether sufficiently large coherent velocities
occur along massive enough filaments for this to be a quantitatively
viable model. Notice that this mechanism should produce enhancements
and decrements of the CMB with equal probability.

The decrement towards PC1643+4631 may not be a unique, or even a
particularly rare object. Another rather similar decrement has 
now been reported. A decrement with a smaller amplitude and on a
smaller angular scale has been mapped at the VLA by \cite{ric97}. 
In this case also a deep {\it ROSAT} HRI observation found no X-ray 
emission. Furthermore, this decrement is also in the direction of an
unusually close QSO pair at $z$ = 2.561. We have applied our analysis 
to this case, for which $\Delta$T$_0 = - 250 \mu$K, assuming $\Theta_c
\sim 15''$ (comparable to the beam size), and the X-ray flux is less than
$2 \times 10^{-14}$ erg cm$^{-2}$s$^{-1}$ at 99.7\% confidence in the 
0.1--2.4 keV band. For a 10 keV cluster these data imply a redshift 
greater than about 0.7. The constraint is weaker than in the case of 
PC1643+4631 because the decrement has smaller amplitude and scale.
Thus PC1643+4631 may be 
the first example of a class of objects which is somehow related
to the background QSO's. As Jones et al noted, the most obvious
relationship might be through gravitational lensing, but to produce
the observed $198''$ separation a filament at $z=2$ would need to
have a mass of $1 \times 10^{16} M_\odot$ 
projected between the two images. This seems
very high, although again it cannot be excluded without detailed simulation
of specific models. 

\section{Conclusion}

Our X-ray limits make it seem unlikely that the decrement towards
PC1643+4631 is due to a distant virialised galaxy cluster.
We cannot definitively exclude the possibility of a hot, 
diffuse and massive cluster containing the two quasars at $z=3.8$,
but such a cluster is difficult to reconcile with present ideas
about structure formation. A supercluster filament seen end-on at
lower redshift may provide a more plausible solution, but then the 
association with the quasar pair is difficult to understand. It seems 
unlikely that this is pure chance since another similar 
decrement has been found also associated with a QSO pair. 
In view of these various problems it is undoubtedly important 
to constrain the microwave spectrum of the CMB decrement, 
to continue searching in other 
frequency ranges, as well as to explore other possible explanations,
for example, topological defects such as textures or strings, exotic
quasar phenomena, or foreground effects of various kinds. 

\section*{ACKNOWLEDGEMENTS}

We are very grateful to Joachim Tr\"umper for supporting this {\it ROSAT} 
observation as part of the final PSPC operation programme, and 
we thank Hans B\"ohringer and Roland Egger 
for valuable suggestions throughout this work. 
We also thank Mike Jones and Richard Saunders for providing 
information on the RT's synthesised beam. RK wishes to 
thank Alastair Edge and Andy Fabian for helpful discussions, 
the Cambridge MRAO group and in particular Anthony Lasenby 
for their very kind hospitality during the completion of this paper 
and the SFB 375 Astro-Particle Physics for additional financial support.


\bsp 

\label{lastpage}

\end{document}